%% file: anycast.tex
\documentclass[11pt,conference]{IEEEtran}
\usepackage{graphicx}

\usepackage{subfigure,url}

\begin{document}
\title{Distributed Overlay Anycast Tables using Space Filling Curves}

\author{\IEEEauthorblockN{Eleni Mykoniati, Lawrence Latif, Raul Landa, Ben Yang, Richard Clegg, David Griffin, Miguel Rio}\IEEEauthorblockN{Department of Electronic and Electrical Engineering, University College London\\Email: \{emykoniati, llatif, rlanda, byang, rclegg, dgriffin, mrio\}@ee.ucl.ac.uk}}
 
\date{}
\maketitle

\begin{abstract}
\input{abstract.tex}
\end{abstract}

\section{Introduction}
\label{section:intro}
\input{introduction.tex}

\section{Background and Related Work}
\label{section:relatedWork}
\input{relatedWork.tex}

\section{Architecture}
\label{sec:architecture}
\input{architecture.tex}

\section{Protocols}
\label{sec:protocols}
\input{protocols.tex}

\section{Evaluation}
\label{section:Eval}
\input{evaluation.tex}

\section{Conclusions and Future Work}
\label{sect:ConcFutureWork}
\input{conclusions.tex}


\bibliographystyle{unsrt}
\bibliography{./anycast.bib}

\end{document}

%% file: abstract.tex
In this paper we present the \emph{Distributed Overlay Anycast Table}, 
a structured 
overlay that implements application-layer anycast, allowing 
the discovery of the closest host that is a member 
of a given group. One application is in locality-aware peer-to-peer networks, 
where peers need to discover low-latency peers  
participating in the distribution of a particular file or stream. 
The DOAT makes use of network delay coordinates and a space filling curve 
to achieve locality-aware routing across the overlay, and Bloom 
filters to aggregate group identifiers. The solution is designed 
to optimise both accuracy and query time, which are essential for 
real-time applications. 
We simulated DOAT using both random and realistic node distributions.
The results show that accuracy is high and query time is low.

%% file: introduction.tex

Anycast is a service that allows a host to discover a \emph{close} host
which is a {\bf member} of a given {\bf group}, where proximity is defined
by a metric like the number of hops or the delay over the IP path. Allowing 
the application to locate nearby object replicas, anycast is 
beneficial to Content Distribution 
Networks\footnote{such as 
\url{http://www.akamai.com} or
\url{http://www.limelight.com}}, 
peer-to-peer data distribution systems and WWW replication architectures.

In anycast systems, hosts register their membership to groups with their
closest anycast-enabled routers. Each group is associated with a distinct
anycast address. Traffic sent to an anycast destination address is routed
to the closest host registered as a member of the corresponding group. It
is up to the routing infrastructure to maintain host group memberships, 
exchange anycast routes and route the data to the closest member host.

However, due to limitations in routing table sizes, addressing schemes and
computational costs, the widespread deployment of anycast at the network
layer (\emph{Network Layer Anycast}, NLA) has been proven problematic, focusing
instead on small groups of application-specific hosts, such as the root DNS
servers.

An alternative to NLA, \emph{Application Layer Anycast} (ALA) bypasses the
addressing and memory constraints in edge and core routers by delegating
anycast routing functions to the hosts themselves. ALA
peers organise themselves in an overlay network, where
links represent peering relationships to exchange routing information and
to forward messages to anycast destination hosts. Unlike NLA, ALA
hosts use the overlay network only to discover anycast destination hosts, not to route
their data. The latter is done separately, over the IP network. The
process of discovering an ALA group member host is called ALA
\emph{querying}.

Although traditional applications can benefit from anycast, the greatest
benefits from an efficient anycast service would be to those applications
that require consistently low-delay interactions,
including live video streaming applications
\cite{coolstreaming},
peer-to-peer virtual environments
\cite{flod} and peer-to-peer networked games \cite{P2PGames}.

In these applications, peers need to discover other peers that are
participating in the distribution of a particular stream, or 
in a given virtual spatial locality, and which are at
the same time able to sustain low delay interactions that allow the system
to maintain high responsiveness and interactivity. With
anycast groups corresponding to a stream or a virtual space,
a fast anycast service can be used by peer selection algorithms to
improve the distribution topology and reduce the
end-to-end path latencies, while at the same time minimising start-up
delay.

Although there have been a number of proposals regarding the implementation
of NLA and ALA, there is still a
need for an anycast architecture focusing on the stringent accuracy and
query resolution time requirements that this new class of real-time
applications presents. To address this, we present \textbf{DOAT
(Distributed Overlay Anycast Table)}, a delay-aware, application-layer
anycast system designed to return accurate results in short time. In
this case, accuracy is associated with the distance, measured as delay over
the IP network path, between the group member discovered by DOAT and the
actually closest group member. Our solution makes use of
three key technologies to implement an ALA service: network
coordinates, space filling curves 
and Bloom filters.

To estimate network delay distance, DOAT nodes use a network coordinate
system to find their location on a multi-dimensional (usually two or three
dimensional) \emph{delay space}. This space is then mapped into a
single-dimensional \emph{DOAT coordinate} using a space-filling curve, in
order to simplify searching operations. This coordinate is used to
construct a locality-aware overlay. Scalability is addressed by
aggregating group memberships using Bloom filters,
and query resolution is accelerated using efficient routing.

%


%% file: relatedWork.tex


Given the great implementation challenges of NLA, there have been a
number of research studies on the subject. Work by Katabi \cite{katabi00framework}
overcomes some of the shortfalls in NLA by using route caching techniques.
Proxies \cite{ballani} are proposed as a means to reduce the size of
routing tables as the number of groups increases.


NLA has no access to metrics such as server load
and available bandwidth, which might be important for certain applications.
Since ALA resides above the network layer, it can monitor both network 
conditions and application-layer metrics. Bhattacharjee
et. al \cite{bhattacharjee1997ala} propose the use of \emph{anycast domain
names} (ADNs) with ADN resolvers maintaining a database of metrics, such as
server load. These metrics are useful only if they accurately reflect the current state
of the network, and thus, techniques are needed to accurately propagate
them. Work presented by Zegura \cite{zegura00applicationlayer} provides a
hybrid server push technique for the maintenance of the metrics database, and
shows that lower response times can be obtained when compared with 
randomly chosen servers. Early work done on ALA \cite{bhattacharjee1997ala, fei98novel} assumes that
groups remain small and have low levels of churn. This last
assumption is rarely justified in peer-to-peer overlays
\cite{stutzbach2006ucp}. \cite{castro03scalable} proposes a solution 
suitable for large groups, assuming however only a small number of groups, 
and loose requirements for accurately routing to the closest group member.

Although DOAT resembles a
Distributed Hash Table (DHT), its function is
fundamentally different.
As in a DHT, DOAT peers form an overlay network where
data (membership entries) can be registered and searched with a given key
(group identifier). However, in DOAT there is no one-to-one mapping 
between the key space and the
overlay nodes.
Group members associated with the same key register in arbitrary locations,
and assignment of group membership data to overlay nodes is based on location,
and not key to node identifier mapping, thus allowing for localised searches.


%% file: architecture.tex
In NLA there is a natural separation between hosts that
make use of the anycast routing capability of the system and the routers
that actually implement it. In the same way, in DOAT 
not all hosts that use the anycast service need to act
as anycast routers. We distinguish between a) \emph{group members},
that register their membership to a group with the anycast system in order
to receive anycast messages, b) \emph{DOAT nodes}, who participate in the
overlay acting as anycast routers, and c) \emph{query senders}, that send
queries to the ALA system in order to discover the closest member
of their group of interest. 

Every group member registers its membership with its closest DOAT node. 
DOAT nodes discover their neighbour nodes and exchange information to 
establish routes to group members. Every query sender is associated
with its closest DOAT node, which will forward its queries into the DOAT overlay, 
following the established routes until the closest group member is found.
The corresponding protocols are detailed in the following section.

%% file: protocols.tex
DOAT uses coordinates to obtain a measure of proximity. In the rest of the paper,
we assume network delay coordinates, however, coordinates
with more rich semantics like \emph{load-aware network coordinates} \cite{LANC}
are also possible. 
In order to minimise query time, DOAT operates on the principle of creating
paths with logarithmically decreasing distances to the destination, a
technique similar to \cite{stoicachord}. This principle underlies the
protocols for establishing connections between the overlay nodes, 
for exchanging routing information and for
forwarding queries for a particular group.

\subsection{Overlay Topology Construction}
\label{subsect:neighbors}
Before connecting to the overlay, a node determines its position
using a distributed coordinate system such as \cite{vivaldi}.  This
allows the node to calculate its coordinates
in a multi-dimensional delay space $\mathcal{X}$ where metric distance
between peers in the space is directly correlated with
network delay between the peers in the network. Thus, short metric
distances imply low delay.

Coordinates in
$\mathcal{X}$ are mapped to a single-dimensional \emph{DOAT coordinate}
which becomes the identifier of the node, used to determine its neighbours in
the overlay.
The DOAT
coordinate has the property that, if two nodes are ``close" in it, 
then they are close in $\mathcal{X}$. 
Note that the opposite is not true: closeness in
$\mathcal{X}$ does not guarantee closeness in the single dimensional DOAT
space. This has the drawback that the closest node in the single-dimensional 
space might not be the closest node in the multi-dimensional
space, but coarser locality information will be preserved.

\begin{figure}[h]
\begin{center}
\includegraphics[width=7cm]{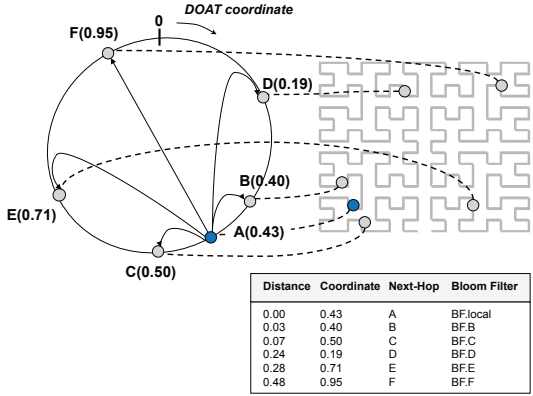}
\caption{DOAT Topology Construction using Space Filling Curve}
\label{fig:doat}
\end{center}
\end{figure}

The mapping from  $\mathcal{X}$ to a 
single-dimensional space is done by
first using a linear transform to map $\mathcal{X}$ coordinates into the
unit square, and
then mapping the unit square to a single-dimensional coordinate using a number of
iterations
of a space filling curve.  The curve used here is the
\emph{H-curve}~\cite{hcurve}, which is known
to have good locality preserving properties.
The obtained single-dimensional coordinate is a wrapping coordinate in the
range of $[0,1)$, that positions the nodes in a ring 
(see figure \ref{fig:doat}).
For the purpose of this paper we will treat $\mathcal{X}$ as a
two-dimensional space, note however that the \emph{H-curve} trivially generalises 
to the multi-dimensional space.

After obtaining its DOAT coordinate, a node establishes peering connections
with other DOAT nodes to exchange routing information. 
Following the principle of logarithmically decreasing
distances, a node first connects to the furthest neighbour (closest to the
opposite point of the ring at $0.5$ distance), then to the two at half
distance ($0.25$) in either direction of the ring, and so on.
This process terminates 
when the immediately closest node is found at each direction. 

To connect to a neighbour $n_2$ at a given distance and direction on the ring, 
a DOAT node $n_1$ calculates a target coordinate for $n_2$. For example,
from $n_1$ at $0.43$, the neighbours at a distance of $0.25$ have target
coordinates of $0.68$ (clockwise) and $0.18$ (anti-clockwise).  
Node $n_1$ then sends a message with the target coordinate $n_2$ to any known
DOAT node $n_3$
, which then forwards the message to its
neighbour that is closest to the target coordinate $n_2$.
This is done recursively, until the message reaches the actual DOAT node
$n_2'$ that is closest to the target coordinate $n_2$. This is the node that
has no neighbour closer to $n_2$ than the node itself.


\subsection{Registering Membership and Updating Routing Tables}
\label{subsect:registering}

When a host becomes a member of a group, it has already
calculated reliably its position in $\mathcal{X}$. It then
sends a registration message to any known DOAT node, which converts the
position of the new group member to the corresponding DOAT coordinate. 
The registration is then forwarded to the DOAT neighbour node 
which is closest to this DOAT coordinate,
until it reaches the DOAT node which is closest to the DOAT coordinate of
the new group member. This node installs the new membership entry at its
local registry.

Each DOAT node maintains a routing table to forward queries for groups
with no members in its local registry. The routing table contains one entry for
the local registry, and one entry for each of its
neighbours. Each entry includes
information on the identity of the next hop (neighbour DOAT node), the
distance to the next hop along the DOAT ring, and the set of groups
reachable through it. Routing entries are sorted in ascending
order of distance to the next hop, with first the entry corresponding
to the local registry. 
Figure \ref{fig:doat} shows the routing table of
the node at DOAT coordinate $0.43$.

The group identifiers cannot be inherently aggregated as it is the case
with, for example, IP addresses. To preserve the scalability of routing
tables and routing update messages, group identifiers are aggregated using
Bloom filters. 

A route announcement from a node $n_1$ to one of its neighbours $n_2$ contains
the groups present in the local registry of $n_1$, and the groups that $n_1$
can reach through other DOAT nodes $n_i$, where the distance between 
$n_1$ and $n_i$ is less than the distance between $n_1$ and $n_2$. In other words, the node
announces all the anycast groups for which there are member hosts in its
local area, to nodes further away.  The area Bloom filter sent to
each neighbour $n_i$ is the aggregate of the Bloom filter that corresponds to the local registry, and the Bloom
filters received from all the neighbours which are closer than $n_i$,
i.e. appearing before $n_i$ in the routing table. 
When a new group is registered with the node $n_1$, the latter sends an update to all its neighbours.
When the node $n_1$ forwards a routing update received from another neighbour $n_2$
then it propagates the update only to these nodes further away than $n_2$.

To reduce the message overhead, a DOAT node may not send routing updates
synchronously with local changes. Instead, a minimum interval may be enforced
between two consecutive updates sent to a neighbour. There is an
accuracy penalty incurred by delaying updating the neighbours
(see Section \ref{subsect:evalperiodic}).
Additionally, instead of sending the entire
Bloom filter every time, a node may send only the difference since the last
update and other compression techniques are possible.

A routing update message also contains the
immediately closest neighbour of the node in both directions. 
These DOAT nodes can be used as alternative neighbours in case the
node fails and does not follow the procedure of gracefully removing itself
from the overlay.

\subsection{Query Forwarding}
\label{subsect:querying}

When a DOAT node receives a query, it will either return the appropriate
group member IP address if the group appears in its local registry, or
forward the query to another DOAT node, according to the Bloom filter
matches on its routing table.

To determine the next hop to forward the query, the DOAT node
will iterate its routing table in increasing distance order, attempting to
match the queried group identifier with the Bloom filters at each step.
The next hop will be the first neighbour whose Bloom filter returns a positive
match with the group identifier. The query is then forwarded to the
associated next hop neighbour. Because the routing table is sorted in ascending order
of distance to the next hop, this is the nearest next hop that has
announced a route for this group. The process is repeated at
each hop until a node with group members in its 
local registry is found.

Because of the neighbour selection and route forwarding mechanisms, the
path the query follows is composed by hops of exponentially decreasing
distances, starting with a maximum of $0.5$ on the DOAT coordinate, i.e.
the maximum network delay between any two DOAT nodes.

\subsection{Node Insertion and Deletion}

The DOAT nodes may be statically appointed hosts, operated by a single
administration. However, the system is designed to work in a peer-to-peer
environment, where any \emph{stable} peer (high uptime, stable network
delay coordinates, considerable storage, bandwidth and processing
resources) can be elected to become a DOAT node
\footnote{We assume that
peers participate in DOAT altruistically or that there is an
incentives mechanism in place \cite{Incentives}.}
(more elaborate methods can be adopted to minimise
churn \cite{godfrey}).

Not every peer becomes a DOAT node and a peer does not decide in isolation
to become a DOAT node. When the load on a DOAT
node increases 
beyond a
locally defined threshold, it chooses the most stable peer in its local
area and sends a
DOAT invite message.

When a node leaves the system
it informs all its neighbours with a message that also includes its
direct neighbours in both directions. This way, all its neighbours can automatically
substitute the removed neighbour with one of its direct neighbours.
Routing updates also carry this information, in case the node fails without
warning. When a node detects that a neighbour has failed, it can
immediately forward queries and routing updates to the 
replacing neighbour of the failed node.

Node removal and re-insertion can be also triggered by network coordinate
drift \cite{wild}, or by changes in the prevailing network congestion and
delay patterns. As the position of a node in the overlay is dependent on
its DOAT coordinate (which itself is calculated from the its network
coordinates), if it moves too far away from its original position, the node
will remove itself from the DOAT and re-insert itself in the new position.
This is necessary because network delay changes that are not reflected in the
DOAT coordinate can reduce the accuracy of the closest group
member discovery (see Section \ref{subsect:evalcoordinates} for an evaluation of the
impact of coordinates accuracy on the DOAT accuracy).

%% file: evaluation.tex
The performance of the DOAT system for a single anycast group is evaluated 
using a discrete event simulator. The number of anycast groups affects 
only the size of the Bloom filters in the routing tables and the 
routing update messages. 

A number of nodes is generated, with each node assigned to a 
two-dimensional coordinate in the network delay space $\mathcal{X}$. 
We used a uniform 
distribution to generate coordinates in the 
range of $[-100,100]$ for each dimension, creating a Euclidean space 
with average delay of around $104$ milliseconds. Three sets of 
$500$, $1000$ and $3000$ nodes were generated. Additionally, 
we used a set of coordinates for $1740$ hosts and an 
average delay of around $145$ milliseconds, generated by applying 
the Vivaldi algorithm \cite{vivaldi} to the delays of the King 
data set \cite{571700}.

Nodes join the DOAT system one by one, 
calculate their DOAT coordinate and connect to neighbours. 
A number of hosts, specified as a percentage over the 
total number of DOAT nodes, register as group members.
This percentage
represents the \emph{density of the group} and affects the 
average distance to the closest group member.

We distinguish between the synchronous and asynchronous update methods. 
In the synchronous update, all group members are registered at 
once, and queries are generated from all the DOAT nodes. 
In the asynchronous update case, each 
time a group member registers, queries are generated from 
$10$\% of the nodes, thus recording 
each of the intermediate states of the system.

For each query, the following metrics are evaluated:
\begin{itemize}
\item \emph{query time}: the sum of the propagation delay along the overlay query forwarding path,
\item \emph{accuracy error}: the difference in distance of the discovered from the actual closest group member; it is is calculated as: \begin{math}
\frac{R - C}{D}
\end{math}, where $R$ is the delay from the querying host to the member host discovered by DOAT, $C$ is the delay from the querying host to the actual closest member host, and $D$ is the average delay in the simulated two-dimensional delay space.
\end{itemize}

For each experiment, we evaluate \emph{overhead} as the number of 
routing messages exchanged per DOAT node, for each registered group 
member. The trade-off between accuracy and overhead is investigated 
in the asynchronous case. An \emph{update interval} is set to 
constrain the frequency of updates a DOAT node can send to any of 
its neighbours. The update interval is specified as multiplies of 
new group member arrival intervals. By reducing the number 
of routing updates overhead is reduced. However, there 
is a penalty in accuracy, as the routing tables are not always up 
to date with routes to reach all the members registered in the system.

Finally, we evaluate the impact of the accuracy of the network delay 
coordinates on the accuracy of the DOAT query results. The experiment 
progresses as in the synchronous case; however, upon evaluating the 
accuracy, we alter the coordinates of the nodes for a random value 
around an average \emph{coordinate offset value}.

\subsection{Query time}
\label{subsect:evalquerytime} 

\begin{figure}[ht!]
\centering{
    \subfigure[Propagation Delay]{
        \includegraphics[width=4cm]{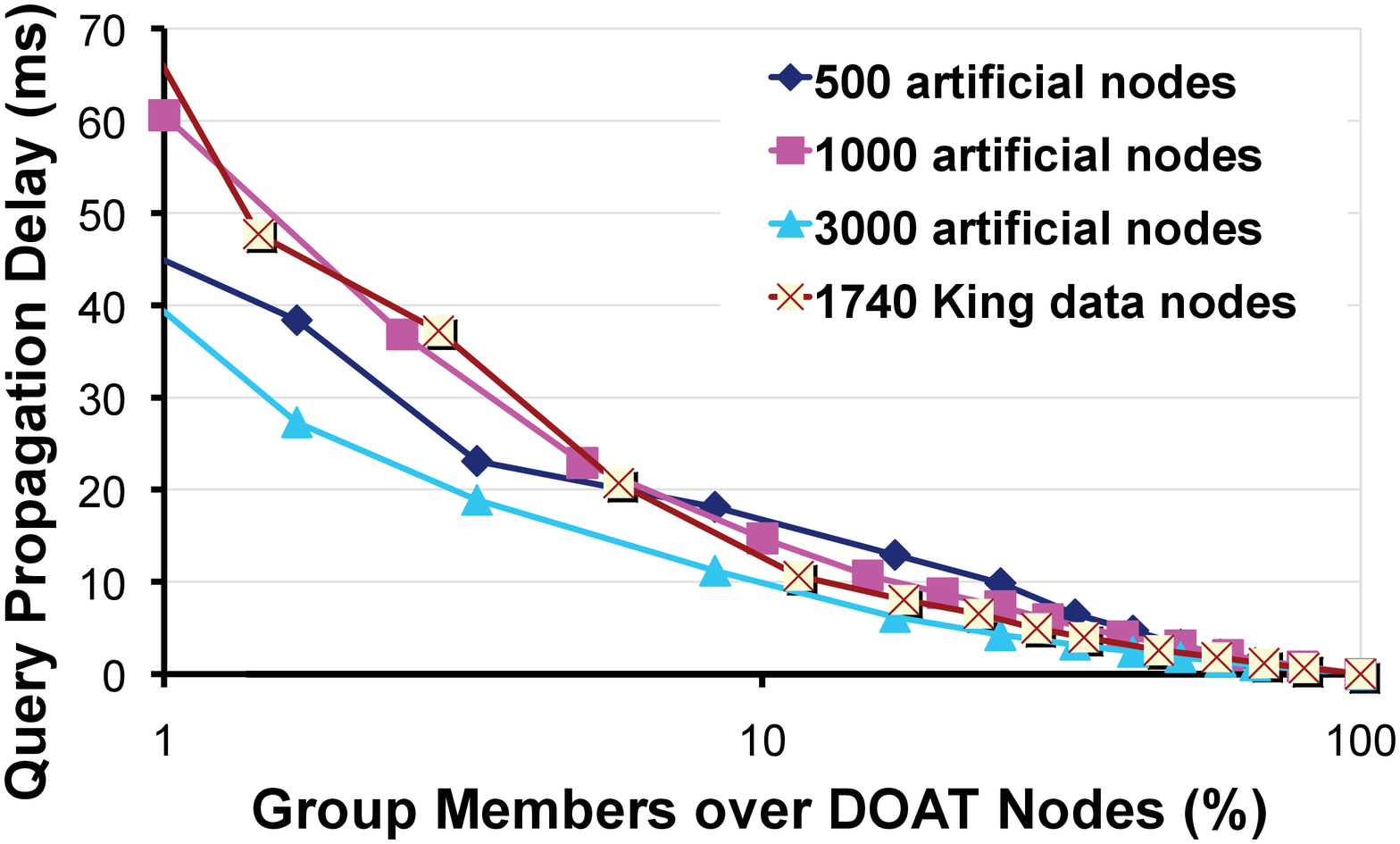}
        \label{fig:qtime_delay}
    }   
    \subfigure[Number of Hops]{
        \includegraphics[width=4cm]{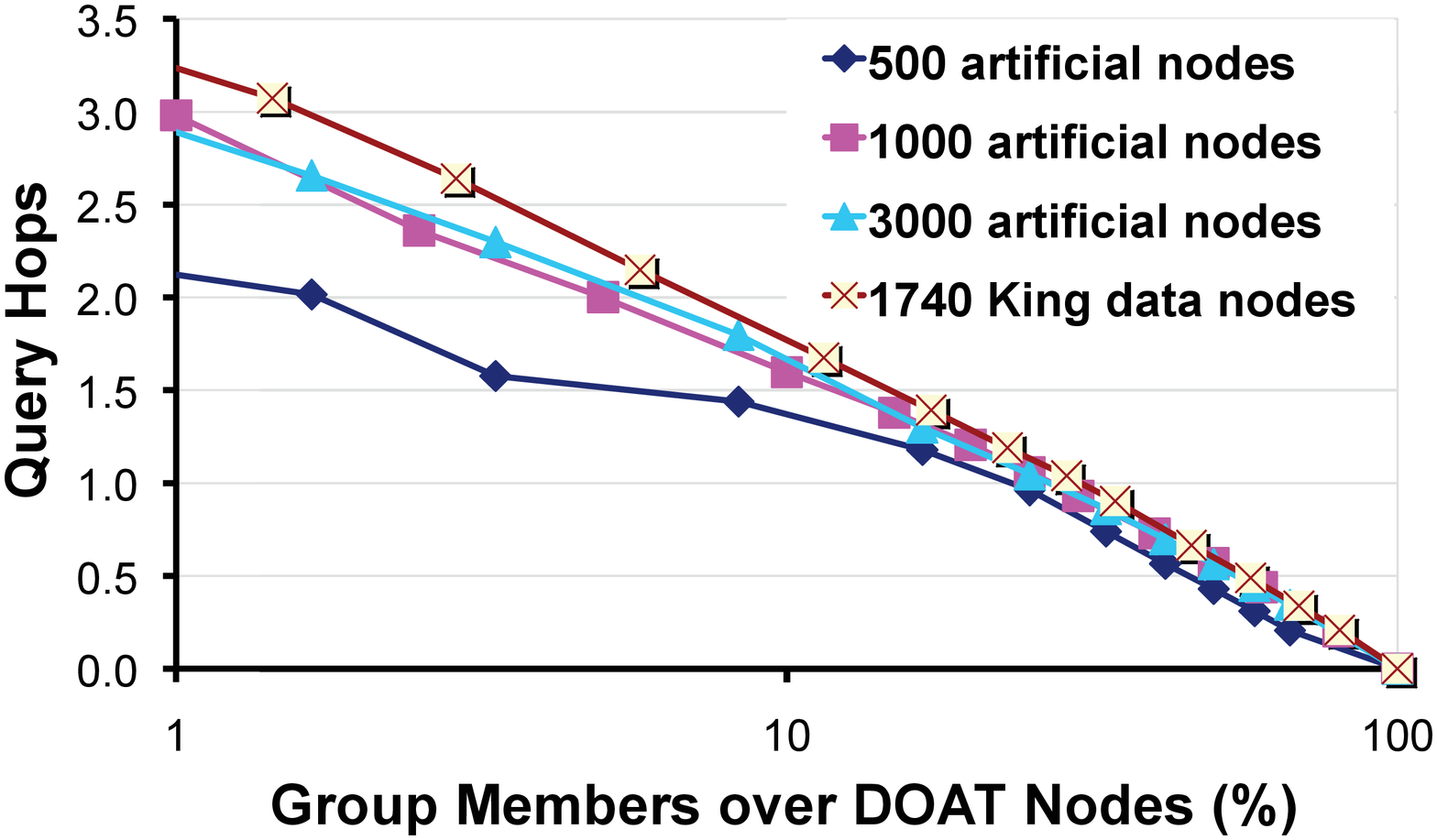}
        \label{fig:qtime_nhops}
    }   
}
\centering{
    \subfigure[Accuracy]{
        \includegraphics[width=4cm]{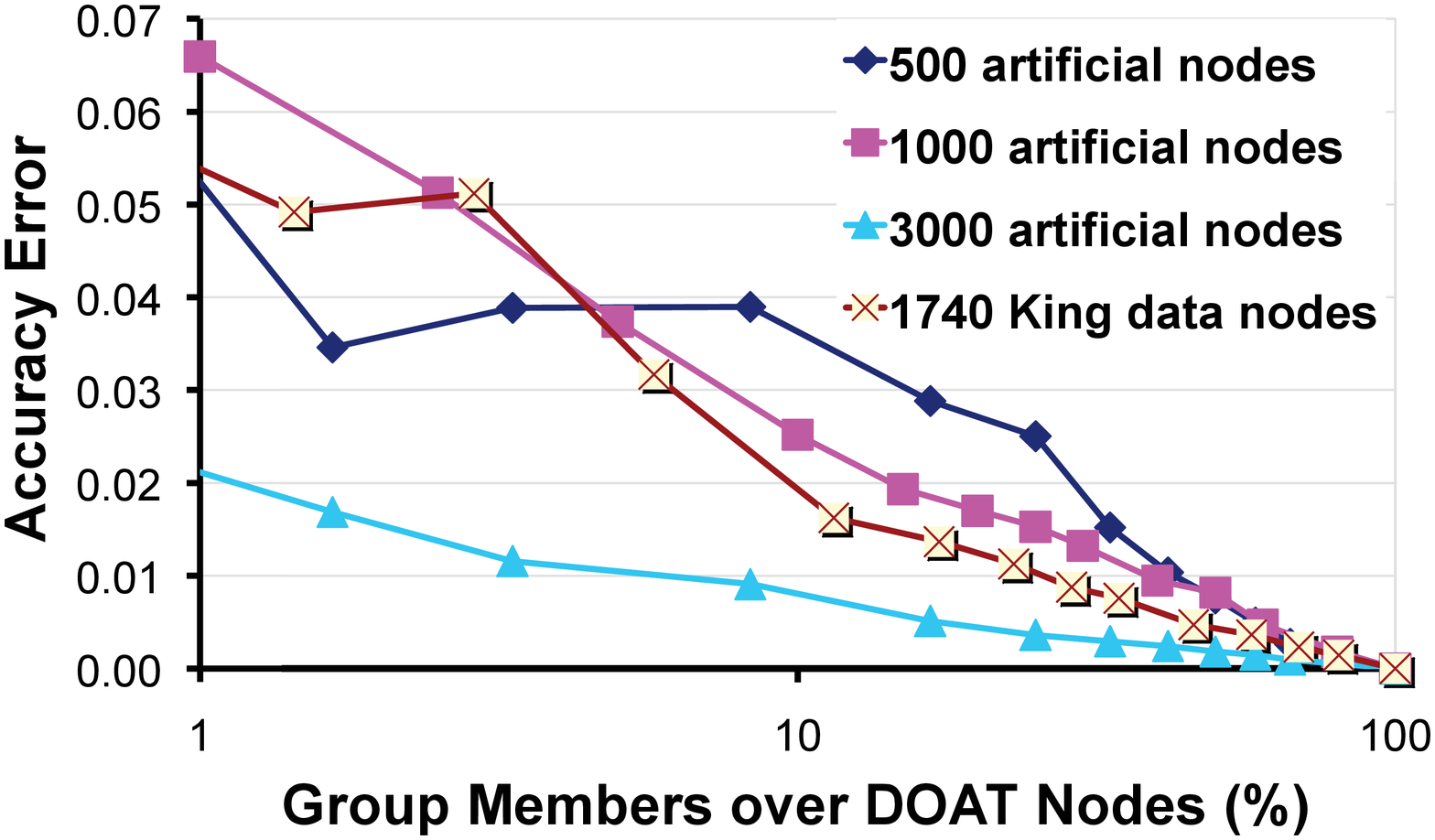}
        \label{fig:accuracy}
    }   
    \subfigure[Overhead]{
        \includegraphics[width=4cm]{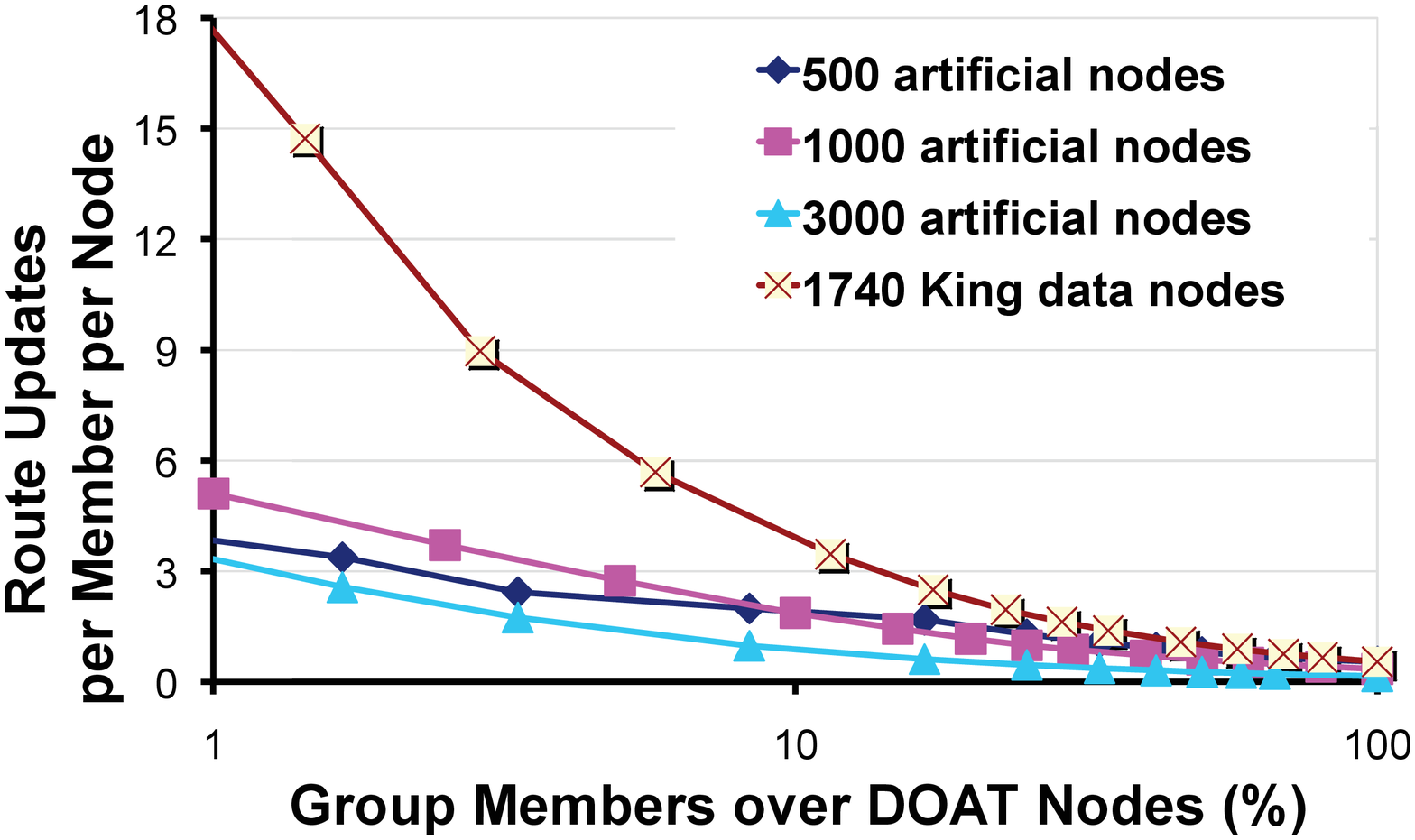}
        \label{fig:overhead}
    }   
}
\caption{Query Time and Accuracy}
\label{fig:queryTime}
\end{figure}

Figures \ref{fig:qtime_delay} and \ref{fig:qtime_nhops} show the average 
query delay and number of hops for the artificial and the King data sets, 
for different values of the ratio of members over DOAT nodes. 
The average query delay to discover the closest member is below the 
average delay between any two nodes, even for the smallest groups, and 
decreases rapidly for larger groups. As the number of DOAT nodes 
increases, the routing tables become more fine-grained, with more 
neighbours in smaller distances. This results in query paths with more 
hops of smaller delays, as can be seen by comparing the values between 
the $500$, $1000$ and $3000$ artificial data sets. Also, comparing the 
uniform delay distribution and the realistic distribution of the King 
data set, we see that the impact is very small in both the delay and 
the number of hops.  

\subsection{Accuracy}

In Figure \ref{fig:accuracy} we see the accuracy for the discovery of the 
closest group member. Even for small groups and small-sized DOAT overlays, 
the accuracy error is below $10$\%. For larger DOAT overlays and/or 
groups the error becomes negligible. One factor that contributes in the 
DOAT accuracy error is the error introduced by the space filling curve 
as there might be neighbours which are actually closer in the network 
delay space $\mathcal{X}$, but appear to be further away in the DOAT 
coordinate space.

\subsection{Overhead}

In Figure \ref{fig:overhead} we can see the number of routing update 
messages exchanged per node for each new member. When there are few members, 
routing updates have to reach all DOAT nodes but as soon as groups 
start growing, messages are only propagated in small regions reducing 
the impact in the overlay.

\subsection{Accuracy and Message Overhead Trade-Off}
\label{subsect:evalperiodic}

\begin{figure}[ht!]
\centering {
    \subfigure[Accuracy]{
        \includegraphics[width=3.95cm]{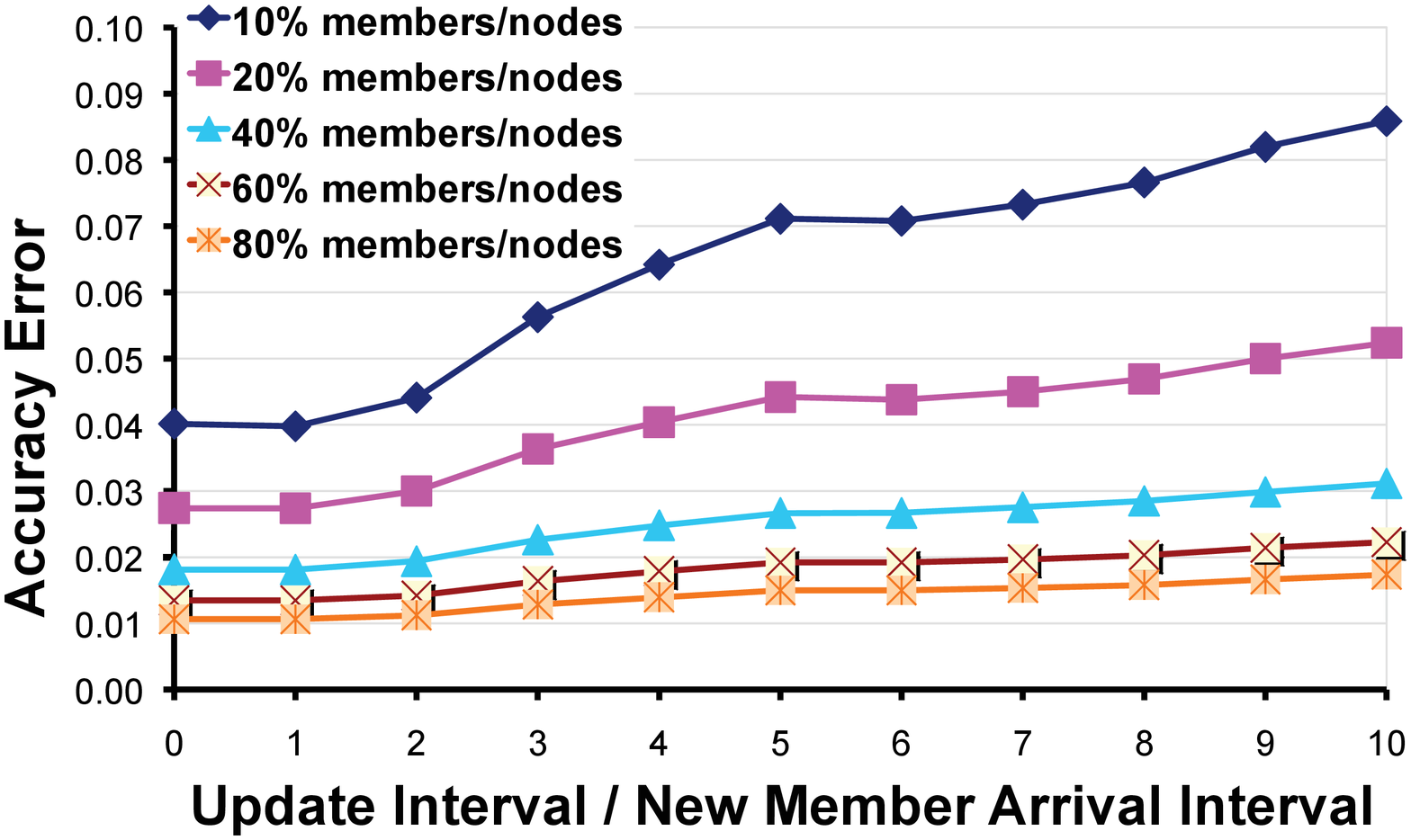}
        \label{fig:AsynchAccuracyReal}
    }
}
\centering {
    \subfigure[Overhead]{
        \includegraphics[width=3.95cm]{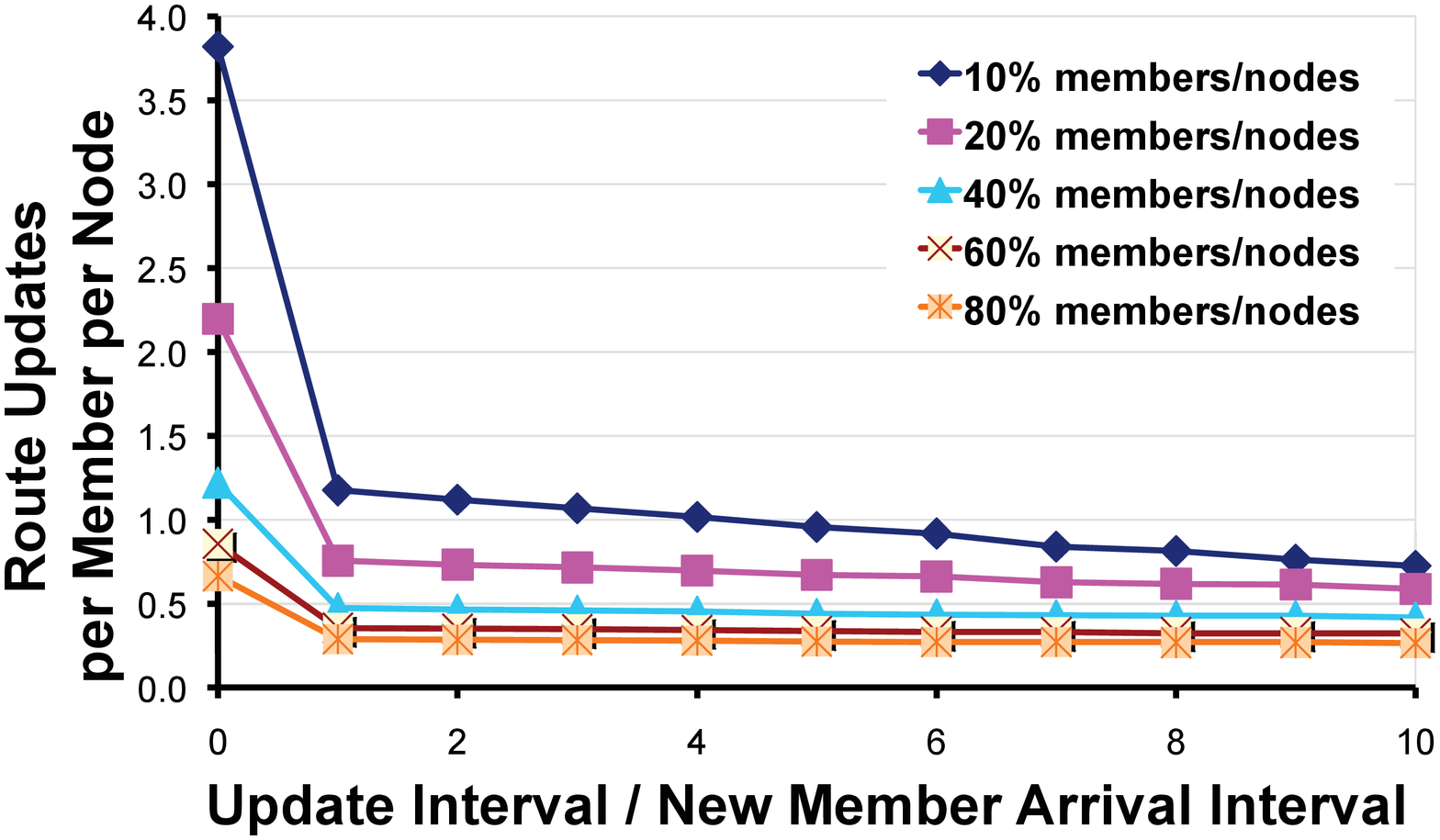}
        \label{fig:AsynchOverheadReal}
    }
}
\caption{Accuracy and Message Overhead Trade-Off}
\label{fig:accuracyOverhead}
\end{figure}

In Figure \ref{fig:accuracyOverhead} we can see the trade-off between 
overhead and accuracy with the King data. Even with a small update interval the overhead 
is significantly reduced, without significantly increasing the accuracy error.

\subsection{Sensitivity to Coordinates Accuracy}
\label{subsect:evalcoordinates}

Network delay coordinates do not represent the network delay with 
accuracy \cite{wild}, or in a static way. To evaluate sensitivity 
to coordinates accuracy, we alter the DOAT node coordinates during the 
course of the experiment by a given offset distance, and we measure the 
impact on the accuracy against the results obtained with static accurate coordinates.

\begin{figure}[ht!]
\begin{center}
\includegraphics[width=4cm]{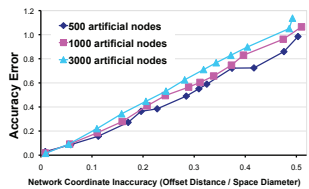}
\caption{Effect of Coordinates Accuracy}
\label{fig:inaccuracy}
\end{center}
\end{figure}

In Figure \ref{fig:inaccuracy} 
one can see a linear relationship between $\mathcal{X}$ accuracy and 
DOAT accuracy. This is considered acceptable, as there are studies in 
the literature \cite{peter} which address the problem of stabilising 
network delay coordinates, enhancing their accuracy on the same time.

%% file: conclusions.tex
We present a structured overlay system that 
implements Application Layer Anycast. The system is designed 
to support a large number of small to very large anycast groups,
with high locality accuracy while minimising the query time. 
We show through simulations in artificial and realistic conditions, 
that 
these objectives are achieved.

One of the future directions is to 
prove the validity of these early results through a 
prototype implementation. This work could then be applied
to large scale systems like the World Wide Web, 
extending HTTP to work with replicated pages. 
This will substantially improve performance and allow small content 
providers to cope with flash crowds and high volumes of hits.

%% file: anycast.bbl
\begin{thebibliography}{10}

\bibitem{coolstreaming}
X.~Zhang, J.~Liu, B.~Li, and Y.-S.P. Yum.
\newblock Coolstreaming/donet: a data-driven overlay network for peer-to-peer
  live media streaming.
\newblock {\em INFOCOM 2005. 24th Annual Joint Conference of the IEEE Computer
  and Communications Societies. In Proc. IEEE}, 3:2102--2111 vol. 3, 13-17
  March 2005.

\bibitem{flod}
Shun-Yun Hu, Ting-Hao Huang, Shao-Chen Chang, Wei-Lun Sung, Jehn-Ruey Jiang,
  and Bing-Yu Chen.
\newblock Flod: A framework for peer-to-peer 3d streaming.
\newblock In {\em INFOCOM 2008. The 27th Conference on Computer Communications.
  IEEE}, pages 1373 --1381, 2008.

\bibitem{P2PGames}
A.~Bharambe, J.~R. Douceur, J.~R. Lorch, T.~Moscibroda, J.~Pang, S.~Seshan, and
  X.~Zhuang.
\newblock Donnybrook: Enabling large-scale, high-speed, peer-to-peer games.
\newblock In {\em Proc. SIGCOMM'08}, August 2008.

\bibitem{katabi00framework}
D.~Katabi and J.~Wroclawski.
\newblock A framework for scalable global {IP}-anycast ({GIA}).
\newblock In {\em {Proc. SIGCOMM'00}}, pages 3--15, 2000.

\bibitem{ballani}
H.~Ballani and P.~Francis.
\newblock Towards a global ip anycast service.
\newblock In {\em Proc. SIGCOMM'05}, August 2005.

\bibitem{bhattacharjee1997ala}
S.~Bhattacharjee, MH~Ammar, EW~Zegura, and V.~Shah.
\newblock {Application-layer anycasting}.
\newblock {\em INFOCOM'97. Sixteenth Annual Joint Conference of the IEEE
  Computer and Communications Societies. In Proc. IEEE}, 3, 1997.

\bibitem{zegura00applicationlayer}
E.~W. Zegura, M.~H. Ammar, Z.~Fei, and S.~Bhattacharjee.
\newblock Application-layer anycasting: a server selection architecture and use
  in a replicated {Web} service.
\newblock {\em IEEE-ACM Trans. Netw.}, 8(4):455--466, 2000.

\bibitem{fei98novel}
Z.~Fei, S.~Bhattacharjee, E.~W. Zegura, and M.~H. Ammar.
\newblock A novel server selection technique for improving the response time of
  a replicated service.
\newblock In {\em {Proc. INFOCOM'98} (2)}, pages 783--791, 1998.

\bibitem{stutzbach2006ucp}
D.~Stutzbach and R.~Rejaie.
\newblock {Understanding churn in peer-to-peer networks}.
\newblock {\em Proc. of the 6th ACM SIGCOMM on Internet measurement}, pages
  189--202, 2006.

\bibitem{castro03scalable}
M.~Castro, P.~Druschel, A.~Kermarrec, and A.~Rowstron.
\newblock Scalable application-level anycast for highly dynamic groups, 2003.

\bibitem{LANC}
N.~Ball and P.~Pietzuch.
\newblock Distributed content delivery using load-aware network coordinates.
\newblock In {\em Proc. of the 3rd International Workshop on Real Overlays and
  Distributed System (ROADS'08), Madrid, Spain}, December 2008.

\bibitem{stoicachord}
I.~Stoica, R.~Morris, D.~Karger, F.~Kaashoek, and H.~Balakrishnan.
\newblock Chord: {A} scalable {Peer-To-Peer} lookup service for internet
  applications.
\newblock In {\em Proc. SIGCOMM'01}, pages 149--160, 2001.

\bibitem{vivaldi}
F.~Dabek, R.~Cox, F.~Kaashoek, and R.~Morris.
\newblock Vivaldi: a decentralized network coordinate system.
\newblock In {\em Proc. SIGCOMM'04}, pages 15--26, New York, NY, USA, 2004.
  ACM.

\bibitem{hcurve}
R.~Niedermeier, K.~Reinhardt, and P.~Sanders.
\newblock Towards optimal locality in mesh-indexings.
\newblock {\em Discrete Applied Mathematics}, 117(1-3):211--237, 2002.

\bibitem{Incentives}
R.~Landa, R.G. Clegg, E.~Mykoniati, D.~Griffin, and M.~Rio.
\newblock A sybilproof indirect reciprocity mechanism for peer-to-peer
  networks.
\newblock In {\em Proc. INFOCOM'09}, April 2009.

\bibitem{godfrey}
P.~B. Godfrey, S.~Shenker, and I.~Stoica.
\newblock Minimizing churn in distributed systems.
\newblock In {\em Proc. of SIGCOMM'06}, pages 147--158. ACM Press, 2006.

\bibitem{wild}
J.~Ledlie, P.~Gardner, and M.~Seltzer.
\newblock Network coordinates in the wild.
\newblock In {\em Proc. of NSDI 2007, Cambridge, MA, April}, 2007.

\bibitem{571700}
K.~P. Gummadi, S.~Saroiu, and S.~D. Gribble.
\newblock King: estimating latency between arbitrary internet end hosts.
\newblock {\em SIGCOMM Comput. Commun. Rev.}, 32(3):11--11, 2002.

\bibitem{peter}
J.~Ledlie, P.~Pietzuch, and M.~Seltzer.
\newblock Stable and accurate network coordinates.
\newblock In {\em Proc. of the 26th International Conference on Distributed
  Computing Systems (ICDCS'06), Lisboa, Portugal}, July 2006.

\end{thebibliography}
